# Large Language Models and User Trust: Consequence of Self-Referential Learning Loop and the Deskilling of Healthcare Professionals

Avishek Choudhury (Ph.D.)[1a*] and Zaira Chaudhry (MD)[1, b]

[1] Industrial and Management Systems Engineering, West Virginia University, Morgantown, WV 26506, US.

[a] avishek.choudhury@mail.wvu.edu;

[b] zsc00004@mix.wvu.edu;

**Corresponding**

Avishek Choudhury (PhD)

Industrial and Management Safety Engineering

West Virginia University

# Large Language Models and User Trust: Consequence of Self-Referential Learning Loop and the Deskilling of Healthcare Professionals

**Abstract**
As the healthcare industry increasingly embraces Large Language Models (LLMs), understanding the intricate dynamics of this integration becomes crucial for maximizing benefits while mitigating potential pitfalls. This paper explores the evolving relationship between clinician trust in LLMs, the transformation of data sources from predominantly human-generated to AI-generated content, and the subsequent impact on the precision of LLMs and clinician competence. One of the primary concerns identified is the potential feedback loop that arises as LLMs become more reliant on their outputs for learning, which may lead to a degradation in output quality and a reduction in clinician skills due to decreased engagement with fundamental diagnostic processes. While theoretical at this stage, this feedback loop poses a significant challenge as the integration of LLMs in healthcare deepens, emphasizing the need for proactive dialogue and strategic measures to ensure the safe and effective use of LLM technology. A key takeaway from our investigation is the critical role of user expertise and the necessity for a discerning approach to trusting and validating LLM outputs. The paper highlights how expert users, particularly clinicians, can leverage LLMs to enhance productivity by offloading routine tasks while maintaining a critical oversight to identify and correct potential inaccuracies in AI-generated content. This balance of trust and skepticism is vital for ensuring that LLMs augment rather than undermine the quality of patient care. Moreover, we delve into the potential risks associated with LLMs' self-referential learning loops and the deskilling of healthcare professionals. The risk of LLMs operating within an echo chamber, where AI-generated content feeds into the learning algorithms, threatens the diversity and quality of the data pool, potentially entrenching biases and reducing the efficacy of LLMs. Concurrently, reliance on LLMs for routine or critical tasks could result in a decline in healthcare providers' diagnostic and thinking skills, particularly affecting the training and development of future professionals. The legal and ethical considerations surrounding the deployment of LLMs in healthcare are also examined. We discuss the medico-legal challenges, including liability in cases of erroneous diagnoses or treatment advice generated by LLMs. The paper references recent legislative efforts, such as The Algorithmic Accountability Act of 2023, as crucial steps toward establishing a framework for the ethical and responsible use of artificial intelligence (AI) based technologies in healthcare. These regulations aim to foster transparency and accountability, ensuring that LLMs are used to enhance patient care while safeguarding against potential misuse. In conclusion, this paper advocates for a balanced and strategic approach to integrating LLMs into healthcare. By emphasizing the importance of maintaining clinician expertise, fostering critical engagement with LLM outputs, and navigating the complex legal and ethical landscape, we can ensure that LLMs serve as valuable tools in enhancing patient care and supporting healthcare professionals. This approach addresses the immediate challenges posed by the integration of LLMs and sets a foundation for their sustainable and responsible use in the future.

# INTRODUCTION

Integration of existing artificial intelligence (AI) models into healthcare—a field where the trust in AI is crucial due to the significant impact of decision-making—is still a work in progress [1]. At the same time, efforts to develop standardized protocols for the deployment of AI in healthcare are underway, yet they have not reached a point of completion [2]. This endeavor is critical for ensuring AI's safe and effective use in healthcare settings. Additionally, the challenge of evaluating AI in healthcare is exacerbated by a lack of comprehensive and standardized metrics [3]. This void is something that researchers and policymakers are actively working to address by creating robust evaluation frameworks that could be applied universally. The regulatory landscape has been focusing on policies around ethical considerations, data privacy, transparency, and patient safety, alongside frameworks that hold AI systems and their developers accountable for the outcomes of their use in patient care [1].

## Advent of generative AI – large language models in healthcare

Despite these ongoing challenges and developments, generative AI like large language models (LLM) is already being deployed in the public sphere [4, 5], utilized by healthcare workers, researchers, and the public for a variety of healthcare-related tasks. Although LLMs have shown promise in medical assessments [6-10], scientific writing, e-healthcare, and patient classification [11-13], its integration marks a shift in paradigm introducing new AI complexities [14-18]. Its rapid and early adoption highlights the critical need for continued discourse ensuring safe and effective integration of LLM into healthcare. Additionally, LLM characters such as —stochasticity, emergent indeterminacy, and lack of consciousness— reinforces the need for cautiousness.

One fundamental aspect of LLMs that prompts special attention is their stochastic paradigm, which means that these models operate based on probabilities and randomness, allowing the model to generate varied outputs for a given input. It exhibits a level of indeterminacy and unpredictability. LLMs can produce different responses under seemingly similar conditions, complicating their reliability. Such LLM behaviors can lead to unexpected results, which, while sometimes beneficial in generating creative solutions or insights, can also pose risks when applied to critical domains like healthcare, where accuracy and predictability are paramount.

Another critical risk characteristic of LLM is the lack of inherent understanding of the context they parse and generate. Despite their ability to produce human-like text, LLMs do not possess consciousness, comprehension, or the ability to discern the truthfulness of their outputs. In other words, LLMs might generate plausible but incorrect content, presenting significant challenges in contexts where the veracity and relevance of information are critical [19, 20].

Approaching AI integration in healthcare with a critical mindset is important. It is crucial for users to have a clear understanding of a technology's actual performance, distinguishing it from the exaggerated expectations set by media hype. These risks underscore the importance of asking the question: are we and our healthcare system ready to integrate LLMs? If yes, is there a policy in place explicitly stating in what capacity it could be used to reduce clinical workload before its dissemination?

## Objective

In this paper, we conceptually investigate the dynamics between clinicians' growing trust in LLMs, the evolving sources of training data, and the resultant implications for both clinician competency and LLM precision over time. Our discussion highlights a potential feedback loop where LLMs, increasingly trained on narrower data sets dominated by their own outputs, may experience a decline in output quality, coinciding with a reduction in user skills. While these phenomena are not yet fully realized, they represent anticipated challenges that coincide with the deeper integration of LLMs into the healthcare domain. We call for preemptive, focused dialogues concerning the integration of LLMs in medical settings, underscoring the importance of maintaining patient safety and the standard of care.

Presently, LLMs are developing at an accelerated pace, heavily reliant on human-generated datasets that are integral to their accuracy and the consequent trust placed in them, particularly in the healthcare sector. This burgeoning dependency, although seemingly beneficial in terms of efficiency and productivity, may lead to an unintended erosion of clinician skills due to the habitual delegation of tasks to AI – as noted in academic context [21, 22]. This trend raises the possibility of an over-reliance on LLM outputs, potentially diminishing the variety and depth of human insights within these models. The risk is a self-perpetuating cycle where LLMs, learning mostly from their creations, could see a degradation in their effectiveness and a narrowing of the breadth of human knowledge they were designed to emulate. Such an outcome would be counterproductive, possibly leading to a decline in both LLM effectiveness and human expertise.

Figure 1 illustrates our core arguments. The first panel reveals a timeline that shows an inverse correlation between clinicians' escalating trust in AI and the preservation of clinical skills over successive time points (T1 to Tn), signaling an increase in AI reliance and a decrease in skill retention. The middle panel demonstrates the shift in training data for LLMs from predominantly human-generated to a growing proportion of AI-generated data, which in turn affects LLM performance and contributes to the feedback loop. The final panel plots LLM accuracy against time, displaying an initial increase as LLMs leverage a mix of data sources. However, upon reaching a tipping point—marked as the self-referential zone—accuracy declines in tandem with the onset of the user deskilling zone, emphasizing the dilemma of increased AI reliance degrading user capabilities. We underscore the need for strategic measures to address these impending challenges in the healthcare sector.

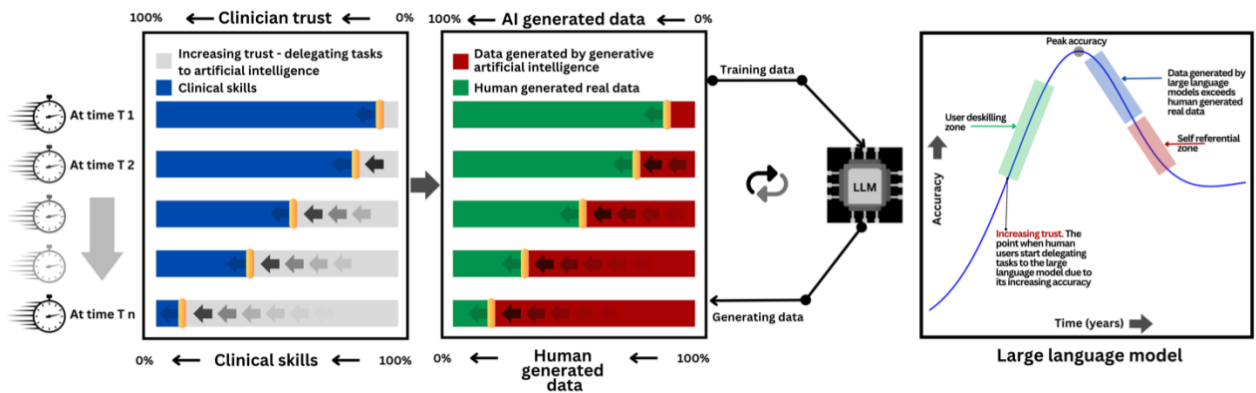

**Figure 1.** The dynamics of user skills, trust, data, and large language models.

## USER EXPERTISE AND TRUST IN LLMS

User trust in LLMs is deeply intertwined with the individual's subject matter expertise and their willingness to engage critically with AI outcomes. Expert users, with a robust understanding of their domain, are more likely to approach LLMs with a discerning mindset and preparedness to review and validate its suggestions. Thus, trust in LLMs can be seen as a spectrum influenced by the user's expertise, and the effort they are willing to invest in ensuring the accuracy of the outcomes.

**User expertise: ability to detect LLM error**

The use of LLMs presents a range of possibilities and challenges that vary depending on the user's expertise and intent, delineating into two primary user categories—subject matter experts and those seeking assistance due to a lack of knowledge.

Subject matter experts (doctors) may employ LLMs to handle routine, time-consuming tasks, enabling them to allocate more time to complex or urgent issues like seeking a second opinion on complex medical diagnoses, or patient triage. They have the advantage of being able to critically evaluate the LLM's output, verify its accuracy (deviation from clinical standards), and make necessary corrections. The expertise of such users acts as a safeguard against potential errors, ensuring that the AI's assistance enhances productivity without introducing risk.

On the other hand, individuals who turn to LLMs due to a lack of expertise in a particular area face a different set of challenges [23]. The ability of LLMs to generate fake but persuasive response further exacerbate the risks making users vulnerable to accepting erroneous information as fact [24]. For instance, a general practitioner faced with a dermatological case, such as an atypical presentation of psoriasis, can utilize an LLM to access detailed diagnostic criteria and treatment protocols. This capability can significantly assist in management of the patient, particularly when the LLM's suggestions are accurate and relevant. However, the inherent risk of LLMs generating incorrect suggestions cannot be overlooked. Such inaccuracies pose a heightened risk to patient safety, especially in scenarios where the clinician may lack the specialized dermatological knowledge required to critically evaluate the validity of the LLM's output [25] – constituting an environment where trust becomes critical.

The crux of the problem lies in the user's ability to verify the accuracy and relevance of the AI-generated content. However, the pivotal consideration here is whether the verification of LLM outcomes by healthcare staff negates the purported reduction in workload. If healthcare professionals are required to meticulously check each AI-generated output for accuracy, the time saved through automation may be offset by the time spent on verification. Maintaining the balance between productivity and accuracy is pivotal. For instance, LLMs can analyze vast datasets to identify patterns or treatment outcomes that may not be immediately apparent to human clinicians, thereby offering insights that can lead to more accurate diagnoses and personalized treatment plans. This capability, even if it requires additional time for verification of AI-generated recommendations, may be deemed a worthy trade-off reducing long-term healthcare costs, and improve care quality. However, this trade-off must be carefully managed to ensure that the pursuit of improved health outcomes does not lead to unsustainable decreases in productivity. Excessive time spent verifying AI recommendations could strain healthcare resources, leading to longer patient wait times and potentially overburdening healthcare staff.

To navigate this trade-off, healthcare systems might adopt strategies such as targeted use of LLMs in high-impact areas where they are most likely to enhance outcomes and the development of systems that prioritize clarity and actionability in their recommendations to minimize verification time. By carefully weighing the benefits of improved patient outcomes against the costs in terms of productivity, healthcare providers can make informed decisions about how best to integrate LLMs into their practices, ensuring that these technologies serve to enhance rather than hinder the delivery of patient care.

**User trust: willingness to review LLM output**

Trust in user engagement with LLMs, particularly in healthcare, is a multifaceted construct influenced by sociotechnical and psychological factors. We acknowledge that user trust in LLMs, in healthcare, can substantially depend on the context. Depending on the stakes (risk) the level of trust required may differ; for instance, LLMs used for diagnosis and treatment recommendations necessitate a higher trust level compared to applications for patient note summarization. Additionally, the degree of autonomy granted to the LLMs, and the extent of clinical oversight are crucial determinants of trust.

Clinicians bring their own norms and expectations to the evaluation of trust in these systems, further complicating the landscape. Individual and cultural perspectives on risk tolerance and acceptance also play pivotal roles. Together, these factors create a complex environment where trust in LLMs is not static but dynamic, varying according to the specific context of use and the interplay of diverse elements. In this section we focus on user willingness to scrutinize LLM output as a precursor to trust.

A user may have the ability and necessary expertise but may not be willing to review LLM generated outcome due to factors including prior trust in the technology or biases. A doctor with high trust (blind trust) in the LLM, might be more inclined to accept its suggestion without extensive further verification [26, 27], exhibiting automation bias [28]. Automation bias, particularly in the context of clinicians' interactions with LLMs can manifests when clinicians exhibit an undue level of trust in the systems, based on past experiences of accuracy and reliability. Blind trust in LLMs can introduce two critical cognitive biases: precautionary [29] and confirmation bias [30], both of which alter clinician behavior in the presence of agreement or disagreement between human judgment and LLM outputs. When LLM recommendations align with a clinician's initial diagnosis or treatment plan (agreement), confirmation bias can be reinforced. Clinicians may overlook or undervalue subsequent information that contradicts the LLM-supported decision, even if this new information is critical to patient care. This confirmation bias can lead to a narrowed diagnostic vision, where alternative diagnoses or treatments are not sufficiently considered. Conversely, in cases where there is a disagreement, precautionary bias can occur. The clinician, having developed a reliance on the LLM due to positive past experiences, might doubt their own expertise and perceive LLM to be the safer alternative for decision-making. Such problem associated with blind trust might persist unchallenged until a point of failure or harm, which can have serious implications in healthcare.

# FUTURE RISK CONSIDERATIONS

As we delve deeper into the dynamics between technology and human expertise, the concepts of the LLM Paradox of Self-Referential Loop and the Risk of Deskilling emerge as pivotal to our discourse. As Figure 1 illustrates the projected trajectory of clinician reliance on LLMs but also hints at the potentially cyclic nature of knowledge and skills within the healthcare industry. Concurrently, the risk of deskilling looms over the horizon, particularly for upcoming generations of healthcare professionals who might become overly reliant on LLMs, possibly at the expense of their diagnostic acumen and critical thinking abilities. This section explores these challenges and the strategies needed to mitigate them. Additionally, this section discusses the LLM accountability concern.

**LLM paradox of self-referential loop (learning from itself)**

In a scenario where LLMs become widely adopted in the healthcare industry for tasks like manuscript writing, educational material creation, clinical text summarization, and risk identification, the possibility of a *self-referential loop* does emerge as a significant concern. This paradox occurs when AI-generated human-like content becomes so widespread that the AI begins to reference its own generated content, potentially leading to an echo chamber effect where original, human-generated insights become diluted or harder to distinguish from AI-generated content. While this problem of a self-referential loop in AI-generated content, particularly in the healthcare industry, has not yet materialized, it represents a likely challenge as generative AI continues to advance and proliferate. The consequence of a self-referential loop in LLMs can lead to several problematic outcomes, including the propagation of biases [31], increased homogeneity in generated data, and ultimately, hindered performance. AI systems learn from the data they are fed, and if these data include biases, the AI is likely to replicate and even amplify these biases in its outputs [32]. In a self-referential loop, the problem becomes compounded. As the AI references its own biased outputs to generate new content, these biases can become more entrenched, making them harder to identify and correct.

The issue of self-referential loops and the potential degradation of information quality are indeed significant concerns; however, when these LLMs, such as Medical Pathways Language Model (Med-PaLM) [33], are specifically fine-tuned and tailored for healthcare applications, the severity of these issues can be mitigated through stringent quality assurance measures. This approach reduces the risk associated with the indiscriminate use of a broader corpus that may contain inaccuracies, outdated information, or irrelevant content. Despite these precautions, the risk of self-referential loops in healthcare contexts can shift towards a different concern: the reinforcement and entrenchment of specific clinical approaches and schools of thought. This occurs as a reflection of the biases present in the curated data sets, which are inherently influenced by the prevailing medical practices, research focus, and therapeutic approaches at the time of data collection.

Addressing this challenge requires a nuanced approach to developing and integrating LLMs technologies into societal frameworks. It involves fostering a symbiotic relationship between human intellect and LLM capabilities, ensuring that AI serves as a tool for augmenting human intellect rather than replacing it. Strategies for maintaining the diversity and quality of training data, including the deliberate inclusion of varied and novel human-generated content, will be critical.

**Risk of deskilling**

As individuals come to rely more on LLMs for routine tasks, such as the synthesis of patient information or the interpretation of medical data, there is a possibility that their skills in these critical areas may diminish over time due to reduced practice [34]. This situation is compounded by the AI's ability to quickly furnish answers to medical inquiries, which might decrease the motivation for in-depth research and learning, consequently affecting the professionals' knowledge depth and critical thinking capabilities.

It is crucial to note that the discussion here does not assert that LLMs will definitively lead to the deskilling of current practitioners in the healthcare sector. These professionals have developed their expertise through extensive experience and rigorous academic training, establishing a solid foundation that is not readily compromised by the integration of AI tools. Instead, the concern is more pronounced for the next generation of healthcare professionals, particularly medical students who might increasingly utilize AI for educational tasks and learning activities where over delegating tasks to AI could attenuate the development of critical analytical skills and a comprehensive understanding of medical concepts, traditionally cultivated through deep engagement with the material [34, 35]. The critical question emerges: *will the ease of generating content with AI stifle the development of creativity and critical thinking in younger generations accustomed to technology providing immediate solutions?*

If future generations of clinicians grow accustomed to AI doing the bulk of diagnostic review and analysis, there is a risk that their own diagnostic skills might not develop as fully. More critically, should they be required to review patient charts manually—due to AI failures—they may find the task daunting, or lack the detailed insight that manual review processes help to cultivate. The crux of the issue lies in ensuring that reliance on technology should not come at the expense of fundamental skills and knowledge. The challenge is to ensure that the deployment of AI technologies complements human abilities without diminishing the need for critical thinking, reasoning, and creativity.

What's needed is to adapt with the paradigm shift – failing to do so can adversely impact healthcare industry. A dual focus on harnessing AI capabilities while enhancing unique human skills is pivotal for advancing patient care in the modern medical landscape. The advent of human-AI collaboration in healthcare, prompts a shift in the skill set emphasis within medical disciplines. The transformation accentuates the value of unique human skills—such as problem-solving, critical thinking, creativity, and fostering patient rapport—over traditional reliance on memory and knowledge base tasks. As LLMs undertake roles in diagnostic assistance, literature synthesis, and treatment optimization, the medical profession should evolve to leverage AI for data-driven insights while prioritizing human-centric skills for patient care. The paradigm shift underscores the growing importance of critical engagement with AI outputs, necessitating medical professionals to adeptly interpret and apply AI-generated information within the complex context of individual patient needs.

## LLM accountability

The integration of LLMs in healthcare introduces medico-legal challenges concerning the allocation and apportionment of liability for outcomes, particularly in instances of negligent diagnoses and treatment. The complexity arises from the interaction between clinicians, healthcare institutions, and AI providers, each contributing differently to the healthcare delivery process.

### Legal Framework and Liability Allocation

In the legal domain, traditional frameworks for medical liability often center on direct human actions, with established principles guiding negligence and malpractice claims. The introduction of LLMs used for diagnostic support or task delegation, complicates these frameworks. Clinicians, operating at the interface of LLM recommendations and patient care, are generally seen as the final decision-makers, thus bearing the primary moral and legal responsibility for the outcomes of those decisions. This perspective is grounded in the principle that clinicians must integrate LLM outputs into a broader clinical judgment context, considering patient-specific factors and adhering to professional standards.

### Shared Liability and AI Providers

However, the role of LLM providers in developing, deploying, and maintaining LLMs introduces questions about shared liability, especially when system errors or deficiencies contribute to adverse outcomes. Determining the extent of LLM provider liability hinges on factors such as the accuracy of the LLM's training data, transparency regarding the system's capabilities and limitations, and the adequacy of user training and support provided.

### Institutional Responsibility

Healthcare institutions also play a critical role in mediating the use of LLMs, responsible for ensuring that these systems are integrated into clinical workflows in a manner that upholds patient safety and complies with regulatory standards. Institutional policies and practices, including the selection of AI tools, clinician training, and oversight mechanisms, are pivotal in mitigating risks associated with LLM use.

### Algorithmic Accountability Act of 2023

The Algorithmic Accountability Act of 2023 and Artificial Intelligence Accountability Act [36, 37] represent a critical legislative step towards ensuring the responsible use of algorithms. The act calls for the creation of standardized procedures and assessment frameworks to evaluate the effectiveness and consequences of these systems, reflecting an understanding of the complex ethical and regulatory challenges posed by AI in decision-making processes, particularly in healthcare. The act is in dialogue with the wider conversation on the ethics of AI, advocating for an approach that emphasizes response-ability—the capacity to respond ethically to the challenges posed by algorithmic decision-making. This perspective is crucial for developing impact assessments and frameworks aimed at promoting fairness and preventing discriminatory practices within algorithmic systems.

The implications of this act on the integration of LLMs in healthcare is profound and ensuring transparency in LLM can further enhance trust in the system. Transparency can allow clinicians to verify errors and review outputs effectively. For example, an LLM providing a diagnostic suggestion would detail the medical literature and patient data informing its analysis, enhancing clinician trust by making the AI's reasoning processes visible and understandable. This transparency combats algorithmic deference by encouraging healthcare professionals to critically assess LLM outputs against their expertise and patient-specific contexts. Moreover, transparency reduces the perceived infallibility of LLMs by highlighting their reliance on input data quality and inherent limitations, promoting a balanced utilization of LLMs as supportive tools in patient care.

## CONCLUSION

It's important to acknowledge that performance of LLMs like ChatGPT today does not guarantee their performance tomorrow. LLMs has the potential to be a substantial boon to the healthcare industry, offering to streamline workflows, enhance the accuracy of patient data processing, and even support diagnostic and treatment planning processes. Its value, however, is contingent upon a systematic and informed integration into healthcare systems. Recognizing that LLMs, like any technology, is fallible is crucial to its successful adoption. Its performance is temporal and will changes as new data is fed to its algorithm. This acknowledgment underpins the necessity for robust oversight mechanisms, ongoing evaluation of AI-driven outputs for accuracy and relevance, and clear guidelines on its role as an assistive tool rather than a standalone decision-maker.

A thoughtful, deliberate approach to integrating generative AI into healthcare can mitigate risks associated with overreliance and deskilling, ensuring that it complements rather than compromises the quality of care. By leveraging AI's strengths and compensating for its limitations through human oversight, healthcare can harness the benefits of this technology to improve outcomes, enhance patient care, and support healthcare professionals in their vital work. Thus, the path forward involves embracing generative AI's potential while remaining vigilant about its limitations, ensuring that its integration enhances rather than diminishes the human element in healthcare.

## CONFLICT OF INTEREST
Authors declare no conflict of interest.

## FUNDING
This study is not funded by any internal or external agency.